\documentclass{aa}
\usepackage{graphicx}
\usepackage{txfonts}  
\newcommand{\Teff}{T_\mathrm{eff}}
\newcommand{\Vmic}{V_\mathrm{mic}}
\newcommand{\Vpec}{V_\mathrm{pec}}
\newcommand{\eps}[1]{\log\varepsilon_\mathrm{#1}}
\newcommand{\kms}{km\,s$^{-1}$}
\begin{document}

\title{Mg, Ba and Eu abundances in thick disk and halo stars 
\thanks{Based on observations collected at the European Southern Observatory, 
Chile, 67.D-0086A,
and the German Spanish Astrono\-mical Center, Calar Alto, Spain}}
\author{L. Mashonkina \inst{1,2,3} \and     
T. Gehren \inst{2} \and C. Travaglio \inst{3,4} \and T. Borkova \inst{5}} 

\offprints{L. Mashonkina, \\ \email{Lyudmila.Mashonkina@ksu.ru}}

\institute{%
Department of Astronomy, Kazan State University, Kremlevskaya 18, 420008 
Kazan 8, Russia 
\and Institut f\"ur Astronomie und Astrophysik der Universit\"at 
M\"unchen, Scheinerstr. 1, 81679 M\"unchen, Germany 
\and Max-Planck-Institut 
f\"ur Astrophysik, Karl-Schwarzschild-Stra{\ss}e 1, 
 D-85740 Garching, Germany
\and Istituto Nazionale di Astrofisica (INAF) - Osservatorio Astronomico di 
Torino, Via Osservatorio 20, 
10025 Pino Torinese (Torino), Italy
\and Institute of Physics, Rostov State University, and Isaac Newton
Institute of Chile Rostov-on-Don Branch, Stachki H. 194, 344090
Rostov-on-Don, Russia
}

\date{Received \today/ Accepted }

\abstract{Our sample of cool dwarf stars from previous papers (Mashonkina \& 
Gehren \cite{euba, eubasr}) is extended in this study including 15 moderately 
metal-deficient stars. The samples of halo and thick disk stars have overlapping 
metallicities with [Fe/H] in the region from $-0.9$ to $-1.5$, and we compare 
chemical properties of these two kinematically different stellar populations 
independent of their metallicity. We present barium, europium and magnesium 
abundances for the new sample of stars. The results are based on NLTE line 
formation obtained in differential model atmosphere analyses of high resolution 
spectra observed mainly using the UVES spectrograph at the VLT of the European 
Southern Observatory. We confirm the overabundance of Eu relative to Mg in halo 
stars as reported in our previous papers. Eight halo stars show [Eu/Mg] values 
between $0.23$ and $0.41$, whereas stars in the thick and thin disk display a 
solar europium to magnesium ratio. The [Eu/Ba] values found in the thick disk 
stars to lie between $0.35$ and $0.57$ suggest that during thick disk formation 
evolved low-mass stars started to enrich the interstellar gas by s-nuclei of Ba, 
and the s-process contribution to barium thus varies from 30\% to 50\%. Based on 
these results, and using the chemical evolution calculations by Travaglio et al. 
(\cite{eu99}), we estimate that the thick disk stellar population formed on a 
timescale between 1.1 to 1.6 Gyr from the beginning of the protogalactic 
collapse. In the halo stars the [Eu/Ba] values are found mostly between $0.40$ 
and $0.67$, which suggests a duration of the halo formation of about 1.5 Gyr. 
For the whole sample of stars we present the even-to-odd Ba isotope ratios as 
determined from hyperfine structure seen in the \ion{Ba}{ii} resonance line 
$\lambda\,4554$. As expected, the solar ratio 82 : 18 (Cameron \cite{cam}) 
adjusts to observations of the \ion{Ba}{ii} lines in the thin disk stars. In our 
halo stars the even-to-odd Ba isotope ratios are close to the pure r-process 
ratio 54 : 46 (Arlandini et al. \cite{rs99}), and in the thick disk stars the 
isotope ratio is around 65 : 35 ($\pm$10\%). Based on these data we deduce for 
thick disk stars the ratio of the s/r-process contribution to barium as 30 : 70 
($\pm$30\%), in agreement with the results obtained from the [Eu/Ba] values. 
}
\maketitle

\section{Introduction}

In this paper we continue our study of element abundances in cool dwarf stars, 
which gives useful information about nucleosynthesis in the Galaxy and also for 
some important parameters of Galactic evolution. In our previous studies 
(Mashonkina \& Gehren \cite{euba, eubasr}, hereafter Paper~I and II) strong 
evidence for a distinct chemical history of the thick and thin disk was found 
from analyses of the Eu/Fe and Eu/Ba ratios: europium is overabundant relative 
to iron and barium in the \emph{thick disk} stars, and there is a step-like 
decrease in the [Eu/Ba] and [Eu/Fe] values at the thick to thin disk 
\emph{transition}. The europium to barium abundance ratio is particularly 
sensitive to whether nucleosynthesis of the heavy elements occurred in the s- or 
r-process. For solar system matter $\eps{Eu,\odot} - \eps{Ba,\odot} = -1.67$ 
(Grevesse et al. \cite{met96}). The contributions of the s- and r-process to the 
solar Ba abundance consist of 81\% and 19\% according to the recent data of 
Arlandini et al. (\cite{rs99}), whereas 94\% of the solar europium originated 
from the r-process. 
This result has been obtained by Arlandini et al. (\cite{rs99}) as the 
best-fit to the solar \emph{main} s-component using stellar AGB models of 1.5 
$M_\odot$ and 3 $M_\odot$ with half solar metallicity. A very similar result for 
Ba and Eu has been obtained by Travaglio et al. (\cite{eu99}) by the 
integration of s-abundances from different generations of AGB stars, i.e. 
considering the whole range of Galactic metallicities.
Thus, the solar abundance ratio of Eu to Ba contributed by 
the r-process relative to the total abundances, [Eu/Ba]$_r$, equals $0.70$. In 
several studies Sneden et al. (\cite{sned96, sned00}), Cowan et al. 
(\cite{Cowan99, Cowan02}), and Hill et al. (\cite{hill}) have presented 
arguments supporting constant relative r-process element abundances during 
the history of 
the Galaxy (at least, where $Z < 70$). Due to the delay in the onset of main 
s-process nucleosynthesis during the thermally pulsing asymptotic giant branch 
(AGB) phase of low-mass stars ($2 \ldots 4 M_\odot$) compared with the 
production of r-nuclei in SNe\,II, the oldest stars in the Galaxy are expected 
to carry a significant Eu overabundance of about $0.7$ dex relative to barium.  
A clear break in the run of [Eu/Ba] values with overall metallicity therefore 
should signal the onset of the contribution to barium coming from AGB stars. Our 
data obtained in Paper~II suggest a dominance of the r-process in heavy element 
production at the epoch of the halo and thick disk formation. Abrupt changes in 
[Eu/Fe] and [Eu/Ba] clearly indicate a hiatus in star formation before the early 
stage of the thin disk developed, when europium enrichment from SN\,II events 
had come to an end, but iron and barium continued to be produced in evolved 
stars of lower mass. 

In this paper we add 15 newly observed stars. For the sample of thick disk stars 
the range of metallicities is now extended to [Fe/H] = $-1.49$, and we first 
determine Eu and Ba abundances in the ``metal-weak thick disk'' stars. The 
sample of metal-poor stars with both Ba and Eu abundance available includes 10 
stars with metal abundances from $-0.90$ to $-1.71$. These new data improve the 
statistical significance of our earlier conclusions. We use the obtained [Eu/Ba] 
abundance ratios to evaluate the ratio of the s/r-process contribution to the 
barium isotopes. Based on chemical evolution calculations of Travaglio et al. 
(\cite{eu99}) we then estimate the timescale for the halo and thick disk 
formation. 

The s/r-process ratio can be found in an independent way from analysis of the 
\ion{Ba}{ii} lines. The idea is based on the fact that the larger the r-process 
contribution is, the larger the fraction of odd isotopes must be, and the 
stronger the hyperfine structure (HFS) broadening of the \ion{Ba}{ii} resonance 
line, $\lambda\,4554$. The even-to-odd Ba isotope ratio is determined from the 
requirement that Ba abundances derived from the \ion{Ba}{ii} subordinate lines 
(which are free of HFS effect) and the resonance line must be equal. In this 
study the even-to-odd Ba isotope ratios are derived for the samples of halo, 
thick and thin disk stars. They reveal a distinction between the different 
Galactic stellar populations. The solar ratio 82 : 18 (Cameron, \cite{cam}) 
adjusts to observations of the \ion{Ba}{ii} lines in \emph{thin disk} stars. A 
mean ratio 65 : 35 ($\pm$10\%) is obtained for \emph{thick disk} stars, whereas 
halo stars are observed with values close to the pure r-process ratio 54 : 46 
(Arlandini et al., \cite{rs99}). 
  
In Paper~I we first reported an overabundance of europium relative to magnesium 
of more than 0.2 dex in two halo stars. The knowledge of Eu/Mg abundance ratios 
in the oldest stars of the Galaxy is of great importance for understanding 
nucleosynthesis in the early Galaxy. One commonly believes that Mg and Eu are 
mainly produced in SN\,II explosions, but the question remains whether  
$\alpha$- and r-process occur in a common site. Theoretical predictions of 
SN\,II element yields show that [$\alpha$/Fe] increases with increasing 
progenitor mass (Arnett \cite{A91}). Most theoretical models of r-process 
nucleosynthesis are based on low mass ($8 \ldots 12 M_\odot$) supernovae 
(Wheeler et al. \cite{rproc}; Tsujimoto \& Shigeyama \cite{Tsujimoto}). However, 
even the lowest mass SN\,II progenitors have very short evolution times of less 
than 20 million years. Thus, if mixing of the interstellar gas in the early 
Galaxy was sufficient, the Eu/Mg ratios in stars born after 20 million years 
from the beginning of protogalactic collapse would be expected to be close to 
solar. Stars born earlier should reveal an underabundance and, certainly, not an 
overabundance of Eu relative to Mg. This problem remained unsolved in Papers~I 
and II because our sample of halo stars was limited to 3 stars. 

Here we determine the [Eu/Mg] abundance ratios for an extended sample of halo 
stars. Eight of the ten stars show an overabundance of europium relative to 
magnesium with a mean value of $0.31 \pm 0.06$ dex,  whereas in both the thick 
and thin disk stars europium follows magnesium. This observational result points 
at different sites for the r-process and $\alpha-$process, respectively, and it 
poses problems on some aspects of Galactic chemical evolution such as mixing of 
the interstellar matter during the halo formation phase, additional sources of 
Galactic magnesium in stars with $M < 8 M_\odot$, and the astrophysical site (or 
sites?) for the r-process. 

Our results are based on high-resolution ($\sim 60\,000$) spectra observed using 
the UVES echelle spectrograph at the ESO VLT2 telescope and the FOCES echelle 
spectrograph at the 2.2 m telescope at Calar Alto Observatory. Stellar 
parameters of the newly included stars are derived with the same methods as  
applied to the remaining stars of our sample which were selected from Fuhrmann's 
(\cite{Fuhr3, Fuhr00}) lists. Effective temperatures $\Teff$ are from Balmer 
line profile fitting, surface gravities $\log g$ from {\sc Hipparcos} parallaxes 
and modelling the line wings of the \ion{Mg}{i}b triplet; metal abundances 
[Fe/H] and microturbulence values $\Vmic$ from profile fitting of \ion{Fe}{ii} 
lines. As in our previous studies barium and europium abundances are based on 
non-local thermodynamical equilibrium (NLTE) line formation for \ion{Ba}{ii} and 
\ion{Eu}{ii}. NLTE magnesium abundances are determined from the \ion{Mg}{i} 
lines using NLTE abundance corrections calculated by Zhao \& Gehren 
(\cite{mgzh01}). 

The paper is organized as follows. Observations and data reduction are 
described in Sect. 2. In Sect. \ref{param} we determine stellar parameters and 
identify the membership of individual stars in particular stellar populations of 
the Galaxy. NLTE magnesium, barium and europium abundances are derived in Sect. 
\ref{abund} and the even-to-odd Ba isotope ratios in Sect.~\ref{isotope}. In 
Sect. \ref{discus} we discuss the element abundance ratios and their 
implications for nucleosynthesis and the evolution of the Galaxy. 

\section{Observations and data reduction}

Our sample of 63 stars from Paper~II is extended 
in this study by including 15 newly observed stars. For the three brightest 
stars, HD\,25329, HD\,148816 and HD\,193901, spectra were observed by Klaus 
Fuhrmann with a resolution of $\sim 60\,000$ and for BD\,$+18^\circ 3423$ by one 
of the authors (TG) at $R \sim 40\,000$ using the fiber optics Cassegrain 
echelle spectrograph FOCES at the 2.2m telescope of the Calar Alto Observatory 
in August and October 2001. The signal-to-noise ratio is $200$ or higher in the 
spectral range $\lambda > 4500$ \AA, but smaller in the blue, where S/N $\sim 
30$ near the Eu line at $\lambda = 4130$ \AA. The data cover an approximate 
spectral range of 4000 - 7000 \AA. 

For metal-poor stars with fainter magnitudes it is necessary to observe with a 
larger telescope because only this guarantees simultaneously high resolution 
\emph{and} high signal-to-noise ratio, both needed to detect and model the 
extremely faint Eu and Ba lines. In April 2001 therefore 14 metal-poor stars 
were observed using the Ultraviolet and Visual Echelle Spectrograph UVES (Dekker 
et al. \cite{UVES}) at the 8m ESO VLT2 telescope on Cerro Paranal. As usual, at 
least two exposures were obtained for each star to keep the influence of hot 
pixels at a minimum. Data extraction followed a full multi-exposure echelle 
analysis originally developed for the FOCES spectrograph and taking advantage of 
data redundancy. This turned out to be much more reliable than standard optimal 
extraction. Finally, the resulting order spectra were rectified with the help of 
spectral continua in neighbouring orders. Typical line profiles are seen in Fig. 
\ref{hd1022}. In spite of the considerable efforts both in observing and in 
extracting the spectra, three of the stars could not be used in our present 
analysis. HD\,140283 is so metal-poor that even with the high S/N ratio the 
\ion{Ba}{ii} $\lambda\,5853$, $\lambda\,6496$ and \ion{Eu}{ii} $\lambda\,4129$ 
lines cannot be extracted from noise. BD\,$-3^\circ 2525$ and CD\,$52^\circ 2174$, 
turn out to show double-lined spectra. 

\section{Stellar parameters} \label{param}

For the three stars marked with an asterisk in Table~\ref{startab}, we use 
stellar parameters determined spectroscopically by Fuhrmann (\cite{Fuhr02b}). 
Effective temperatures $\Teff$ are found from Balmer line profile fitting, 
surface gravities $\log g$ from analysis of the line wings of the Mg~Ib triplet; 
metal abundances [Fe/H] and microturbulence velocities $\Vmic$ are derived from 
\ion{Fe}{ii} line profile fitting. For the remaining stars listed in 
Table~\ref{startab} stellar parameters are determined using the same methods,  
with $\log g$ calculated from the {\sc Hipparcos} parallaxes. Our analyses are 
all based on the same type of model atmospheres. In stellar parameter 
determinations we use the MAFAGS line-blanketed LTE model atmospheres generated 
and discussed by Fuhrmann et al. (\cite{Fuhr1}). 

{\it Effective temperatures:~} Theoretical H$_\alpha$ and H$_\beta$ line 
profiles are calculated according to the description given by Fuhrmann et al. 
(\cite{Fuhr93}). For each star the difference of effective temperatures obtained 
from H$_\alpha$ and H$_\beta$ is less than 100~K, and for 12 stars the mean 
value $\Delta\Teff$(H$_\alpha$ - H$_\beta$) = $-20 \pm 60$~K. We adopt the 
average value of $\Teff$(H$_\alpha$) and $\Teff$(H$_\beta$) as final effective 
temperature and estimate a statistical error of $\Teff \sim 60$~K. 

{\it Surface gravities:~} Nearly all stars investigated here have {\sc 
Hipparcos} parallaxes (ESA \cite{ESA}) with an accuracy $\sigma(\pi)/\pi < 0.2$ 
(the only exception being BD$-4^\circ 3208$ with $\sigma(\pi)/\pi = 0.27$). The 
well-known relation between $g$, stellar mass $M$, radius $R$ and $\Teff$ is 
used to calculate $\log g$\,({\sc Hip}), where square brackets denote the 
logarithmic ratio with respect to the solar value,  
$$ [g] = [M] + 4[\Teff] + 0.4 ({\rm M_{bol}}-{\rm M_{bol,\odot}}) \quad . $$ 
Here, ${\rm M_{bol}}$ is the absolute bolometric magnitude, with bolometric 
corrections taken from Alonso et al. (\cite{BC}). For the Sun, an absolute 
visual magnitude ${\rm M_{V,\odot}} = 4.83$ and bolometric correction BC$_\odot 
= -0.12$ (Allen \cite{Allen}) are adopted. The mass is obtained by interpolating 
in the ${\rm M_{bol}}$ vs. $\log T_{\rm eff}$ diagram between the 
$\alpha$-element enhanced isochrones of 12~Gyr and 16~Gyr calculated by 
VandenBerg et al. (\cite{VdB2000}). The internal precision of the mass is 
estimated to be better than $\pm0.05 M_\odot$. 

The error of the surface gravity $\log g$\,({\sc Hip}) derived from {\sc 
Hipparcos} parallaxes is dominated by the error of $\pi$, where $\sigma(\pi)/\pi 
= 0.20$ transforms to $\sigma(\log g) = \pm 0.16$~dex. Column 4 of 
Table~\ref{startab} lists $\log g$\,({\sc Hip}) and the errors obtained from 
adding the squared errors of parallax and mass. 
\begin{table}[htbp] 
\caption{ Stellar parameters of the new sample} 
\label{startab}  \tabcolsep1.2mm  
\begin{tabular}{rcccccc}   
\noalign{\smallskip} \hline \noalign{\smallskip} 
 HD/BD & $\Teff$ & Mass & \multicolumn{2}{c}{$\log g$} & [Fe/H] & $\Vmic$  \\
       &  [K]    &[$M_\odot$]& ({\sc Hip}) & (\ion{Mg}{i})   & &  [\kms]   \\
\noalign{\smallskip} \hline \noalign{\smallskip}
25329$^*$ & 4800 &   &               & 4.66 & --1.84 & 0.6  \\
29907\,~  & 5500 & 0.7~ & 4.64$\pm$0.06 & 4.67 & --1.55 & 0.6  \\
31128\,~  & 5980 & 0.8~ & 4.49$\pm$0.07 & 4.42 & --1.49 & 1.2  \\
34328\,~  & 5955 & 0.8~ & 4.54$\pm$0.07 & 4.44 & --1.61 & 1.2  \\
59392\,~  & 6010 & 0.85 & 4.02$\pm$0.15 & 3.90 & --1.59 & 1.4  \\
74000\,~  & 6225 & 0.85 & 4.16$\pm$0.16 & 4.28 & --2.00 & 1.4  \\
97320\,~  & 6110 & 0.8~ & 4.27$\pm$0.05 & 4.27 & --1.18 & 1.4  \\
99383\,~  & 6100 & 0.8~ & 4.22$\pm$0.12 & 4.37 & --1.54 & 1.4  \\
102200\,~ & 6115 & 0.8~ & 4.20$\pm$0.08 & 4.24 & --1.24 & 1.4  \\
122196\,~ & 6000 & 0.9~ & 3.99$\pm$0.12 & 3.94 & --1.71 & 1.5  \\ 
148816$^*$ & 5880 &  &               & 4.07 & --0.78 & 1.2  \\ 
193901$^*$ & 5780 &  &               & 4.46 & --1.08 & 0.9  \\ 
298986\,~ & 6130 & 0.8~ & 4.30$\pm$0.16 & 4.22 & --1.34 & 1.4  \\
$-4^\circ$3208\,~ & 6280 & 0.9~ & 4.08$\pm$0.24 & 4.03 & --2.23 & 1.6  \\
$18^\circ$3423\,~ & 6070 & 0.8~ & 4.28$\pm$0.15 & 4.14 & --0.90 & 1.4  \\
\noalign{\smallskip} \hline \noalign{\smallskip}
\end{tabular}
  
$^*$ {\footnotesize stellar parameters from Fuhrmann (2002b)}
\end{table}  %

In addition, the surface gravities were determined by an independent method 
based on the \ion{Mg}{i} $\lambda\,5172$ and $\lambda\,5183$ line wing fitting. 
First the magnesium abundance is obtained from the weaker \ion{Mg}{i} $\lambda\,
5528$ and $\lambda\,5711$ lines, which are not strong enough to develop 
significant line wings. Then the damping of the \ion{Mg}{i}b lines is used as an 
indicator of surface gravity. LTE was assumed both in line wing fitting and in 
deriving the abundance. Since we use a differential analysis with respect to the 
Sun, atomic parameters of the \ion{Mg}{i} lines are improved empirically from 
analyses of the \emph{solar} line profiles (Kurucz et al. \cite{Atlas}) on the 
base of the MAFAGS solar model atmosphere and assuming $\Vmic = 1$ \kms. The 
values of $\log gf\varepsilon_\odot$ and $\log C_6$ obtained for the solar 
spectrum are given in Table~\ref{data}. We note that using the meteoritic value 
$\eps{Mg} = 7.58$ (Grevesse et al., \cite{met96}) as the solar photospheric Mg 
abundance the empirically determined values of $\log gf$ coincide within $0.05$ 
dex with the corresponding oscillator strengths from Opacity Project 
calculations (Butler et al. \cite{OP}). For $\lambda\,5172$ and $\lambda\,5183$ 
the $\log C_6$ values in Table~\ref{data} are in agreement with $\log C_6 = 
-30.69$ based on Anstee \& O'Mara's (\cite{anstee}) calculations. For $\lambda\,
5528$ our value of $\log C_6$ is smaller by $0.15$ dex compared with $\log C_6 = 
-29.94$ based on the calculations of Barklem \& O'Mara (\cite{bark97}). We refer 
to abundances on the usual scale where $\eps{H} = 12$. 

\begin{table}[htbp] 
\caption{Atomic data obtained from solar line profile fitting for the \ion{Mg}{i} 
and \ion{Fe}{ii} lines used in stellar parameter determinations} 
\label{data}  
\tabcolsep1.2mm  
\begin{center}
\begin{tabular}{cccc}   
\noalign{\smallskip} \hline \noalign{\smallskip} 
$\lambda$ (\AA) & E$_\mathrm{low}$(eV) & $\log gf\varepsilon_\odot$ & $\log C_6$ \\ 
\noalign{\smallskip} \hline  \noalign{\smallskip} 
\multicolumn{4}{c}{\ion{Mg}{i} lines} \\ 
5528.41 & 4.33 & 7.03 & --30.10 \\ 
5711.09 & 4.33 & 5.86 & --30.18 \\ 
5172.70 & 2.70 & 7.13 & --30.69 \\ 
5183.62 & 2.70 & 7.35 & --30.75 \\ 
\multicolumn{4}{c}{\ion{Fe}{ii} lines} 
\\ 4923.93 & 2.88 & 6.00 & --31.91 \\ 
5018.45 & 2.88 & 6.22 & --32.11 \\ 
5264.81 & 3.22 & 4.43 & --32.19 \\ 
5425.26 & 3.19 & 4.24 & --32.19 \\ 
5325.56 & 3.21 & 4.30 & --32.19 \\ 
5234.63 & 3.21 & 5.19 & --31.89 \\ 
5197.58 & 3.22 & 5.19 & --31.89 \\ 
6456.38 & 3.89 & 5.40 & --32.18 \\ 
6247.56 & 3.87 & 5.16 & --32.18 \\ 
\noalign{\smallskip} \hline \noalign{\smallskip} 
\end{tabular}  
\end{center}
\end{table}       

Column 5 of Table~\ref{startab} lists $\log g$\,(\ion{Mg}{i}). It is obvious  
that for most of the stars surface gravities obtained from the {\sc Hipparcos} 
parallax and \ion{Mg}{i}b lines agree within error bars of $\log g$\,({\sc 
Hip}). For HD\,34328 and HD\,99383 the difference $\log g$\,({\sc Hip}) $-\log 
g$\,(\ion{Mg}{i}) only slightly exceeds $1\,\sigma$. For 12 stars this 
difference equals, on average, $0.02 \pm 0.09$~dex. Based on the data for $\sim 
100$ stars Fuhrmann (\cite{Fuhr00}) has found a systematic deviation of the 
surface gravities based on {\sc Hipparcos} parallaxes from those based on 
\ion{Mg}{i}b spectroscopy of $0.02 \pm 0.04$~dex. Since both methods give very 
similar results, we adopt $\log g$\,({\sc Hip}) as final values for the 
investigated stars. 
 
{\it Fe abundance and microturbulence:~} From comparison of the spectra of 
typical program stars with the solar spectrum 9 unblended \ion{Fe}{ii} lines 
were selected to determine the Fe abundance and microturbulence velocities 
$\Vmic$ by the requirement that the derived [Fe/H] abundances should not depend 
on line strength. We assume LTE in the analyses of the \ion{Fe}{ii} lines as 
verified in our recent NLTE calculations for \ion{Fe}{i} and {\sc ii} (Gehren et 
al. \cite{fe_nlte}) which have shown neglegible NLTE effects for \ion{Fe}{ii}. 
Van der Waals damping constants $C_6$ were taken from Kurucz's (\cite{kur92}) 
line list. Similar to the \ion{Mg}{i} lines the values $\log 
gf\varepsilon_\odot$ for \ion{Fe}{ii} were obtained from solar line profile 
fitting. They are given in Table~\ref{data}. Using $\eps{Fe} = 7.51$ as solar 
iron abundance the obtained values $\log gf$ for 6 \ion{Fe}{ii} lines coincide 
within $0.11$ dex with the corresponding oscillator strengths from Raassen \& 
Uylings (\cite{fe_gf}), and the mean difference equals $0.06 \pm 0.06$. 

For each star the line-to-line scatter of [Fe/H] does not exceed $0.13$~dex, and 
the mean value is calculated with a mean square error of no more than 
$0.04$~dex. We estimate that the error of $\Vmic$ is about $0.1$ \kms. The 
uncertainties in $\Teff$ (60~K) and $\log g$ (0.1~dex) correspond to errors of 
$\leq 0.02$~dex and $0.05$~dex in [Fe/H]. Altogether, the error of [Fe/H] is of 
the order of $0.1$~dex except for the most distant star BD$-4^\circ 3208$ with a 
maximum uncertainty in $\log g$ roughly twice as large. 

{\it Membership in Galactic stellar population:~} All the stars have [Fe/H] 
$\leq -0.90$, and they can be expected \emph{not} to belong to the thin disk. We 
follow Fuhrmann (\cite{Fuhr00, Fuhr02}) and use the stellar kinematics to 
discriminate between halo and thick disk stars. The kinematic data were taken 
from the {\sc Hipparcos} catalogue (ESA \cite{ESA}) and the catalogue of radial 
velocities of Barbier-Brossat \& Figon (\cite{Vrot}). For three stars with 
halo-type metal abundances, HD\,31128 ([Fe/H] = $-1.49$), HD\,97320 ([Fe/H] = 
$-1.18$) and HD\,102200 ([Fe/H] = $-1.24$), the peculiar space velocities $\Vpec 
= 117$ \kms, $85$ \kms and $157$ \kms, respectively, favour a thick disk 
membership. This assumption is supported by their small $W$ space velocity 
components. Possibly, these stars represent the so-called ``metal-weak 
thick-disk'' discussed in the literature by Norris et al. (\cite{norris}) and 
Fuhrmann (\cite{Fuhr02}). The lowest metal abundance of $-1.84$ in our sample of 
thick disk stars was found by Fuhrmann (\cite{Fuhr02}) for HD\,25329. For the 
remaining stars $\Vpec$ is between $224$ \kms and $440$ \kms, and we refer to 
them as halo stars. 

\section{Element abundances} \label{abund}

For each star line-blanketed LTE model atmospheres have been generated using the 
MAFAGS code at given values of $\Teff$, $\log g$, [Fe/H] (Table~\ref{startab}) 
and [$\alpha$/Fe], where [$\alpha$/Fe] is the relative abundance of the most 
abundant $\alpha$-process elements O, Mg and Si, of which the last two 
contribute in significant amounts to the electron pressure in cool stellar 
atmospheres. We assume that oxygen and silicon abundances follow magnesium and 
adopt [$\alpha$/Fe] = [Mg/Fe]. 

\subsection{Mg abundances}

Magnesium abundances were derived in this study from profile fitting of the  
\ion{Mg}{i} $\lambda\,5528$ and $\lambda\,5711$ lines, originally assuming LTE. 
NLTE abundance corrections $\Delta_\mathrm{NLTE}$ from Zhao \& Gehren 
(\cite{mgzh01}) were then added to obtain NLTE abundances. NLTE effects tend to 
weaken the \ion{Mg}{i} lines due to photoionization. For both lines this implies 
\emph{positive} corrections $\Delta_\mathrm{NLTE}$ which are only slightly 
different by value. Relative to the solar values $\Delta_\mathrm{NLTE} 
(\lambda\,5528) = 0.02$ dex and $\Delta_{\rm NLTE}(\lambda\,5711) = 0.05$ dex. 
NLTE abundance corrections for the investigated stars are therefore mostly 
between $0.06$~dex and $0.10$~dex. For 11 stars the difference between NLTE 
abundances obtained from $\lambda\,5528$ and $\lambda\,5711$ does not exceed 
$0.07$~dex, with a mean value of $0.04 \pm 0.02$~dex. Only for HD\,99383 the 
corresponding difference of $0.10$~dex is slightly larger. The final Mg 
abundance is obtained as the average value. The [\ion{Mg}{i}/\ion{Fe}{ii}] 
ratios are presented in Table~\ref{elabund}. 

\begin{table}[htbp] 
\caption{Element abundances of the new sample} \label{elabund}  
\begin{tabular}{rrrrrl}   
\noalign{\smallskip} \hline \noalign{\smallskip} 
 HD/BD & [Fe/H] & [Mg/Fe] & [Ba/Fe] & [Eu/Fe] & Note \\
\noalign{\smallskip} \hline \noalign{\smallskip}
25329$^*$ & --1.84 & 0.42 &   0.33 & 0.24~~ & N-rich \\
29907\,~  &    --1.55 & 0.29 & --0.02 & 0.63~~ & \\
31128\,~  &    --1.49 & 0.34 &   0.02 & 0.44~~ & \\
34328\,~  &    --1.61 & 0.38 &   0.20 & 0.27~~ & ?  \\
59392\,~  &    --1.59 & 0.27 &   0.19 & 0.68~~ & \\
74000\,~  &    --2.00 & 0.35 &   0.24 & 0.16~: & N-rich \\
97320\,~  &    --1.18 & 0.36 &   0.02 & 0.40~~ & \\
99383\,~  &    --1.54 & 0.37 &   0.07 & 0.40~~ & SB \\
102200\,~ &    --1.24 & 0.30 & --0.01 & 0.57~~ & \\
122196\,~ &    --1.71 & 0.16 & --0.07 & 0.24~~ & \\ 
148816$^*$ & --0.78 & 0.41 & --0.13 & 0.30~~ & \\ 
193901$^*$ & --1.08 & 0.18 & --0.04 & 0.44~~ & \\ 
298986\,~ &    --1.34 & 0.16 & --0.03 & 0.54~~ & \\
$-4^\circ$3208\,~ & --2.23 & 0.34 & --0.14 & --~~~ & \\
$18^\circ$3423\,~ & --0.90 & 0.12 & --0.02 & 0.42~: & \\
\noalign{\smallskip} \hline \noalign{\smallskip}
\end{tabular}  

$^*$ {\footnotesize [Fe/H] and [Mg/Fe] are from Fuhrmann (2002b)}
\end{table}  %

\subsection{Ba and Eu abundances}

\begin{figure}
\resizebox{88mm}{!}{\includegraphics{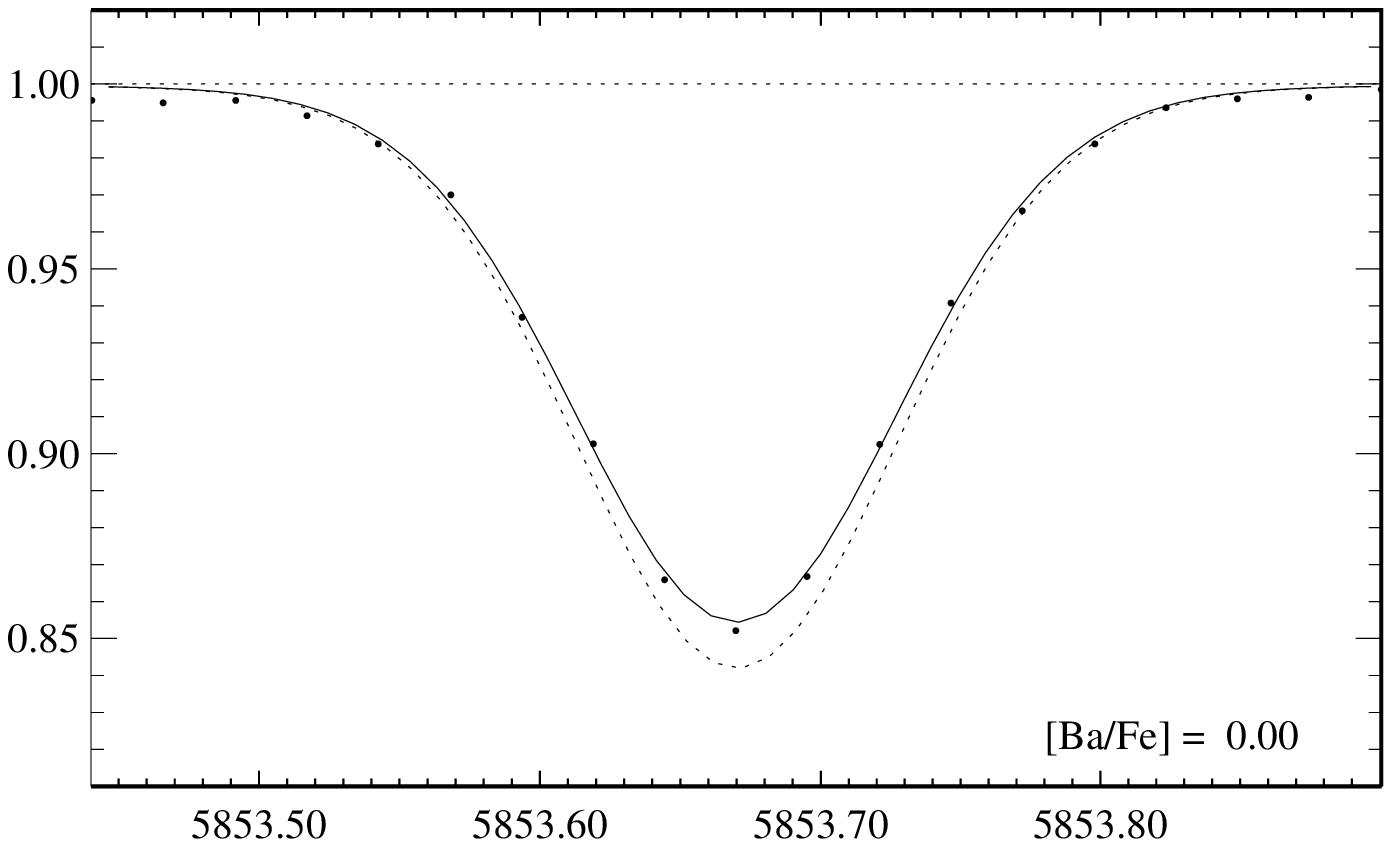}} \vspace{-0mm} 
\resizebox{88mm}{!}{\includegraphics{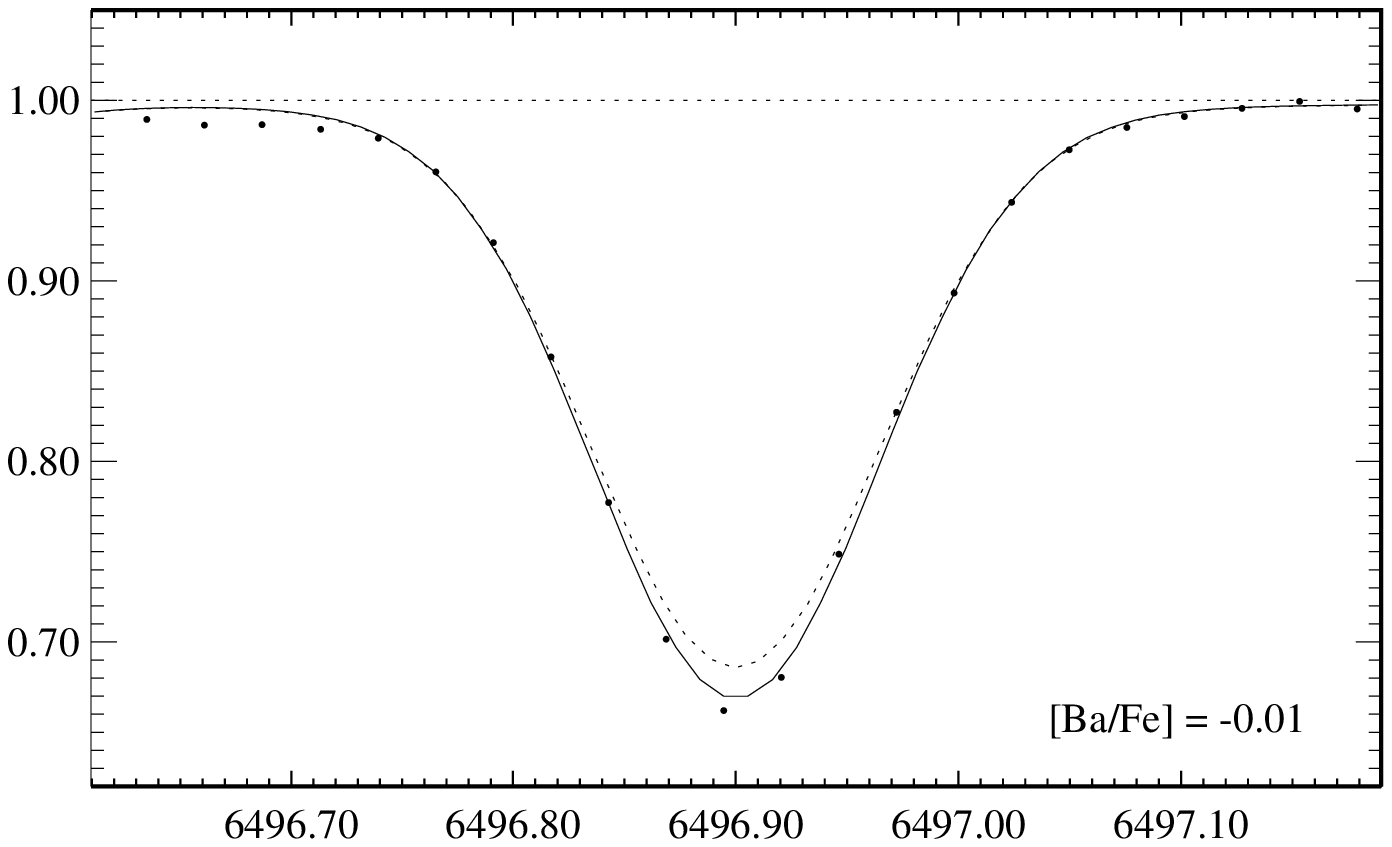}} \vspace{-0mm} 
\resizebox{88mm}{!}{\includegraphics{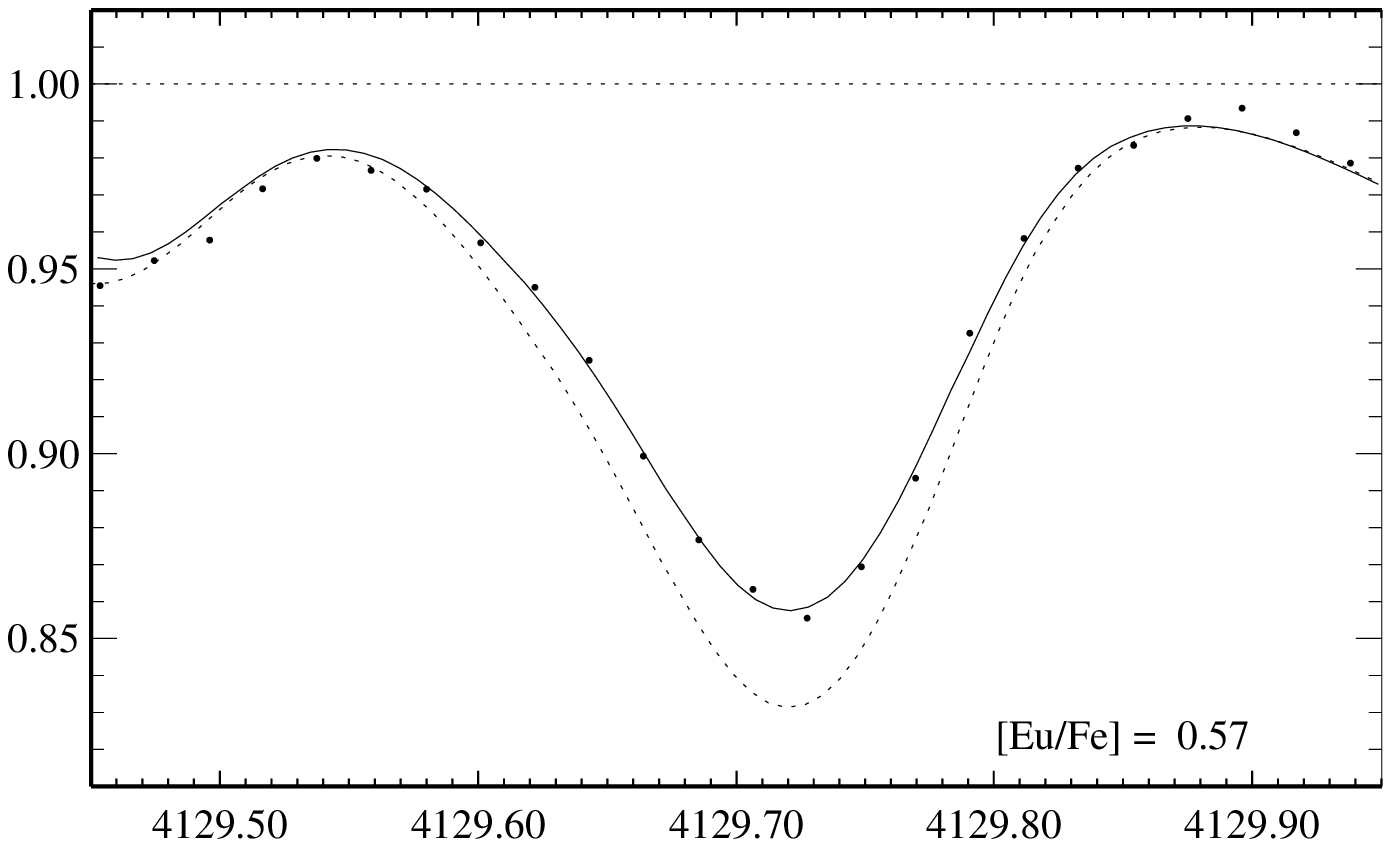}} \vspace{-3mm} 
\caption[]{Synthetic NLTE (continuous line) and LTE (dotted line) flux profiles
of Ba and Eu lines compared with the observed UVES spectra (bold dots) 
of HD\,102200 ([Fe/H] = $-1.24$)}
\label{hd1022}
\end{figure}

We use the same method as in Papers I and II to derive Ba and Eu abundances for 
the stars. The synthetic line profiles are computed using the departure 
coefficients of the \ion{Ba}{ii} and \ion{Eu}{ii} levels from the code NONLTE3 
(Sakhibullin \cite{Sa}) and the LTE assumption for other atoms. The line list is 
extracted from Kurucz' (\cite{Kur94}) compilation, and it includes all the 
relevant atomic and molecular lines. A differential analysis with respect to the 
Sun is performed. Solar barium and europium abundances, $\eps{Ba,\odot} = 2.21$ 
and $\eps{Eu,\odot} = 0.53$, and van der Waals damping constants $C_6$ for the 
\ion{Ba}{ii} and \ion{Eu}{ii} lines were determined in Paper I from solar line 
profile fitting. The methods of NLTE calculations for \ion{Ba}{ii} and 
\ion{Eu}{ii} were developed earlier (Mashonkina \& Bikmaev \cite{Mash96}; 
Mashonkina et al. \cite{Mash99}; Mashonkina \cite{MLeu}; Paper I). Some examples 
of the \ion{Ba}{ii} and \ion{Eu}{ii} line profile fitting are given in Fig. 
\ref{hd1022}. 

Altogether, uncertainties in $\Teff$ (60 K), $\log g$ (0.1 dex) and $\Vmic$ (0.1 
\kms) cause Ba abundance errors up to $0.08$ dex in the range of metal abundances 
between $-0.9$ and $-1.5$ and up to $0.05$ dex for the more metal-poor stars. For 
europium abundances the corresponding values are $0.06$ dex and $0.05$ dex. However, 
the ratios [\ion{Ba}{ii}/\ion{Fe}{ii}] and [\ion{Eu}{ii}/\ion{Fe}{ii}] are much 
less affected by possible errors of stellar parameters. Test calculations for 
BD$+18^\circ 3423$ ([Fe/H] = $-0.90$) have shown that varying $\log g$ by $0.15$ dex 
leads to $\Delta$[Ba/Fe] $= 0.04$ dex and $\Delta$[Eu/Fe] $= 0.02$ dex. The effect 
is even smaller ($\sim 0.01 \ldots 0.02$ dex) at [Fe/H] $\sim -1.5$. 

{\it Barium} abundances are obtained from the two \ion{Ba}{ii} subordinate lines, 
$\lambda\,5853$ and $\lambda\,6496$, nearly free of hyperfine structure. In three 
stars $\lambda\,6496$ could not be used because of blends with telluric lines. In 
the spectrum of BD$-4^\circ 3208$, $\lambda\,5853$ is too weak and could not be 
extracted from noise. 

As discussed in a previous paper (Mashonkina et al. \cite{Mash99}), NLTE effects 
for \ion{Ba}{ii} depend on the Ba abundance, which correlates with the overall 
metal abundance of the atmospheric model. Thus, NLTE leads to stronger  
\ion{Ba}{ii} lines compared with LTE in stars of normal or moderately deficient 
metal abundances, but changes to the opposite effect in very metal-poor stars. 
$\Delta_\mathrm{NLTE}$ changes its sign at [Fe/H] between $-1.2$ and $-1.9$ 
depending on $\Teff$ and $\log g$. As most stars of our new sample have [Fe/H] 
in this metallicity range, NLTE abundance corrections have different signs in 
different stars: $\Delta_\mathrm{NLTE}(\lambda\,6496)$ varies from $-0.14$ dex 
to $0.15$ dex and $\Delta_\mathrm{NLTE}(\lambda\,5853)$ from $-0.01$ dex to 
$0.10$ dex. For 3 stars NLTE effects are opposite for $\lambda\,6496$ and 
$\lambda\,5853$ (see, for example, Fig. \ref{hd1022}). For 12 stars the mean 
value of the difference between NLTE abundances derived from $\lambda\,6496$ and 
$\lambda\,5853$ equals $-0.01 \pm 0.02$ dex, while under the assumption of LTE 
Ba abundances from the first line are systematically overestimated relative to 
$\eps{LTE}(\lambda\,5853)$, with a mean difference of $0.05 \pm 0.05$ dex. 
Taking into account the uncertainties of the stellar parameters we estimate the 
total statistical error of [Ba/Fe] to be $\pm 0.05$ dex. 

{\it Europium} abundances have been derived for 14 stars of our new sample from 
the \ion{Eu}{ii} line, $\lambda\,4129$. In the spectrum of BD$-4^\circ 3208$ 
this line was again too weak, and could not be extracted from noise; therefore, 
only an upper limit of $\sim 0.4$ dex for the [Eu/Fe] value was estimated. The 
$\lambda\,4129$ line is located in a crowded spectral range and the uncertainty 
in determining the local continuum may cause Eu abundance errors up to $0.05$ 
dex for the spectra observed with UVES, and up to $0.10$ dex for the FOCES 
spectra. 

Atomic data for hyperfine structure and isotopic shift were described in detail 
in Paper I for $\lambda\,4129$. As discussed there, NLTE effects weaken the 
\ion{Eu}{ii} $\lambda\,4129$ line compared with the LTE case, and NLTE abundance 
corrections are therefore positive. For our stars $\Delta_\mathrm{NLTE}$ ranges 
from $0.05$ dex to $0.11$ dex. 

\begin{figure} 
\resizebox{88mm}{!}{\includegraphics{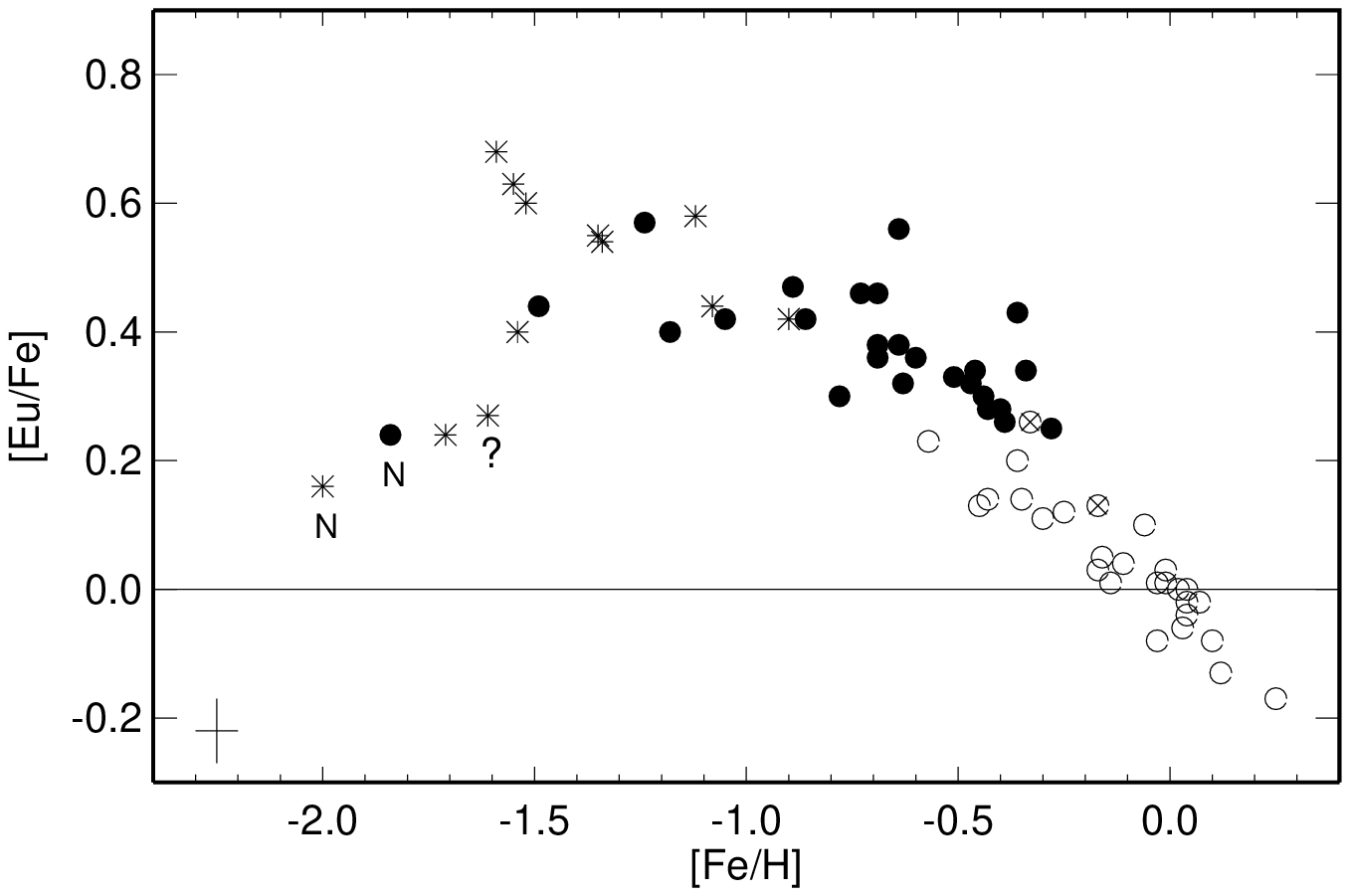}} 
\resizebox{88mm}{!}{\includegraphics{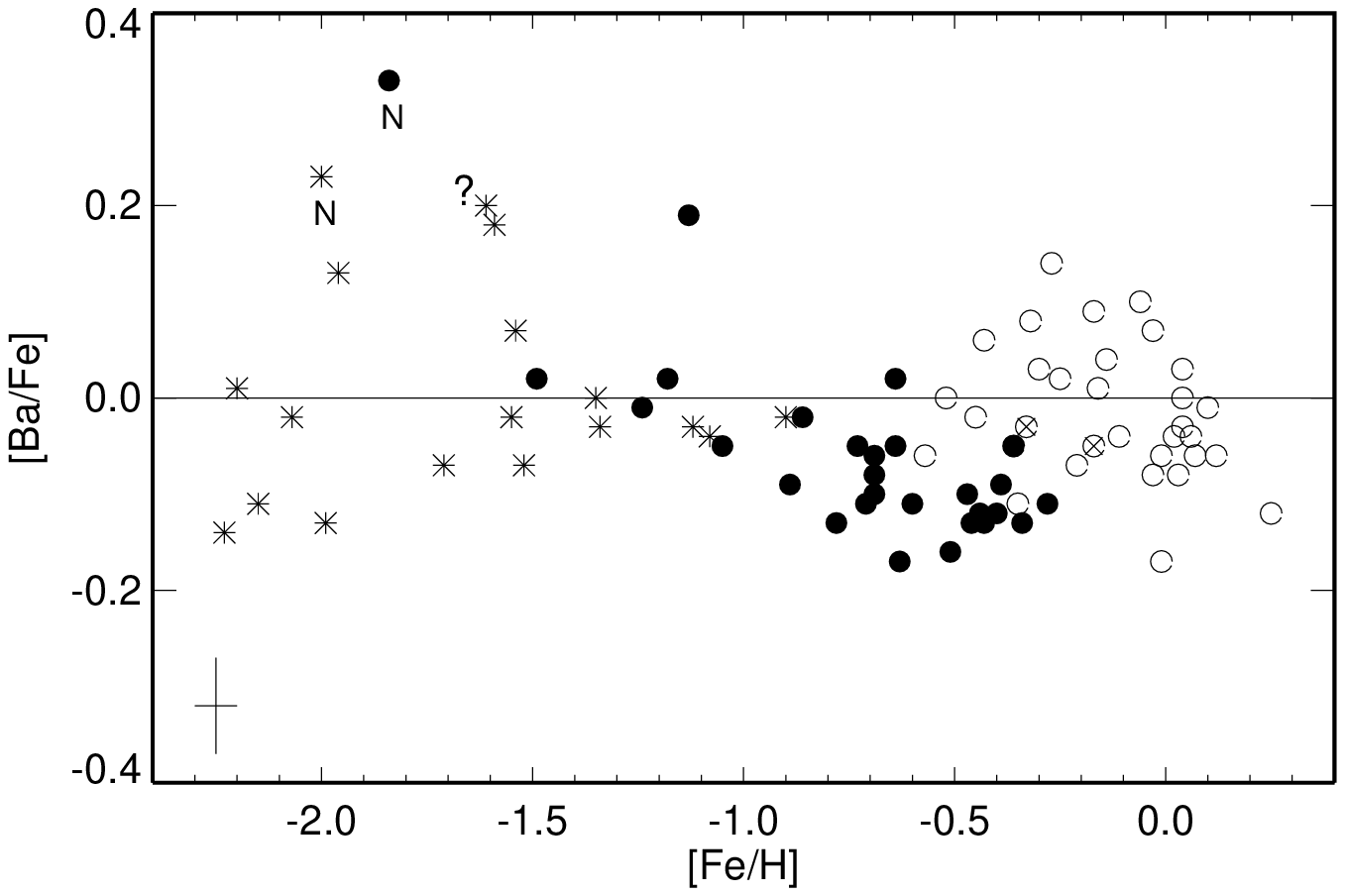}} 
\caption[]{The run of [Eu/Fe] 
(top panel) and [Ba/Fe] (bottom panel) with [Fe/H]. Symbols correspond to thin 
disk (open circles), thick disk (filled circles), and halo stars (asterisks). 
The two stars indicated by a cross in an open circle are transition stars 
between thin and thick disk according to Fuhrmann (1998). The N-rich stars are 
marked by ``N'' and HD\,34328 with similar chemistry by ``?''. Error bars are 
indicated at the lower left} \label{eu_fe} 
\end{figure}

The results are presented in Table \ref{elabund}. [Ba/Fe] and [Eu/Fe] values for 
the whole sample of stars are shown in Fig. \ref{eu_fe}. We comment on a few 
stars which reveal different element abundances compared with other stars of 
similar metal abundances. 

\underline{HD\,25329 and HD\,74000} are known as nitrogen-rich stars (Carbon et 
al. \cite{nrich}) with [N/Fe] = 0.5 and 0.9, respectively. For the typical halo 
star HD\,103095 the same authors give [N/Fe] = --1.0. Our data show for both 
stars overabundances of barium (Table \ref{elabund}) and strontium (preliminary 
result) relative to iron of more than 0.2 dex, whereas in the remaining 
metal-poor stars Ba and Sr are slightly \emph{underabundant}. The most 
surprising result is presented by their low Eu abundances, with [Eu/Fe] = $0.24$ 
and $0.16$ for HD\,25329 and HD\,74000, respectively. The [Eu/Ba] values 
($-0.09$ and $-0.08$) are close to solar, and they suggest a dominance of the 
s-process contribution to barium in contrast to other halo and thick disk stars, 
which show a significant contribution of the r-process. Simultaneously, both 
stars have an overabundance of Mg relative to iron of $\sim 0.4$ dex, typical 
for the halo and the thick disk. High N and s-process element abundances could 
be explained by contamination of the surface layers with products of 
nucleosynthesis during the AGB phase of an evolved primary component. However, 
neither our spectra nor any publication suggests that these stars are binaries. 
Nevertheless, we suppose that N-rich stars do not represent \emph{normal} 
chemical evolution of Galactic matter, and we exclude HD\,25329 and HD\,74000 
from further analysis. 

\underline{HD\,34328} was not studied with respect to its N abundance but the 
[Ba/Fe], [Sr/Fe], [Eu/Fe] and [Mg/Fe] values in this star turn out to be very 
close to the corresponding values in the two N-rich stars. Most probably, 
similar nucleosynthesis processes are responsible for the chemical peculiarity 
of N-rich stars and also HD\,34328, and we exclude the latter from further 
analysis. 

\underline{HD\,122196} shows low overabundances not only of europium but also of 
magnesium relative to iron of the order of 0.2 dex. Kinematically ($\Vpec = 224$ 
\kms), according to metal abundance ([Fe/H] = $-1.71$) and our age estimate 
($\sim 12$ Gyr) it is a member of the halo population. 

\underline{HD\,99383} is noted by Nissen et al. (\cite{nis02}) as spectroscopic 
binary. Our UVES spectrum of this star does not show double lines. However, we 
have already mentioned slightly larger uncertainties of the surface gravity and 
Mg abundance compared with the other stars. Thus larger errors of Ba and Eu 
abundances may be expected. 

The important results of our previous studies (Papers I and II) were based on 
element abundance ratios related to the Eu abundance. In this paper our sample 
of halo stars with Eu abundances available is extended to 10 stars with [Fe/H] 
ranging between $-0.90$ and $-1.71$ and for the sample of thick disk stars the 
range of metallicity is extended down to $-1.49$. Our new data improve the  
significance of the results obtained in Papers I and II. We summarize them as 
follows. 
\begin{itemize}
\item 
In halo stars europium is overabundant relative to iron with [Eu/Fe] values 
between $0.40$ and $0.67$ (HD\,122196 is the only exception) and there is a 
marginal trend of increasing [Eu/Fe] with decreasing metal abundance. The 
[Ba/Fe] values are mostly between $0.12$ and $-0.14$. The relatively large value 
[Ba/Fe] = $0.18$ found for HD\,59392 is accompanied by high Eu abundance with 
[Eu/Fe] = $0.67$ which may reflect a local inhomogeneity of the interstellar 
matter. 
\item 
In thick disk stars europium is overabundant relative to iron  with a clear 
decline of the [Eu/Fe] abundance ratios from about $0.50$ at [Fe/H] = $-1.5$ to 
$0.25$ at [Fe/H] = $-0.3$. The decline of [Ba/Fe] values with increasing metal 
abundance  noted in Paper II for thick disk stars becomes more evident for our 
extended sample of stars. 
\item
In the region of overlapping metallicities (from $-0.9$ to $-1.5$) both the 
[Ba/Fe] and [Eu/Fe] values do \emph{not} reveal a distinction between the halo 
and thick disk stars, in contrast to the thick-to-thin disk transition where a 
step-like change of the [Eu/Fe] and [Ba/Fe] values occurs. 
\end{itemize}

\section{Even-to-odd Ba isotope ratios} \label{isotope}

Previously (Mashonkina et al. \cite{Mash99}) we suggested a direct method for 
the evaluation of even-to-odd Ba isotope ratios from the \ion{Ba}{ii} resonance 
line $\lambda\,4554$, provided that the Ba \emph{abundance} can be determined 
from the \ion{Ba}{ii} subordinate lines. The resonance line of the odd isotopes 
has several HFS components and this leads to an additional broadening of the 
line. HFS components of each odd isotope appear in two groups shifted relative 
to the resonance line of even isotopes by 18 and $-34$ m\AA, respectively. Our 
test calculations for both solar metal abundance and metal-poor stars have shown 
that the 3-component simplification suggested by Rutten (\cite{Rut}) is 
sufficiently precise and we accept it in this study. The relative strengths of 
the components depend on the even-to-odd isotopic ratio. Oscillator strengths of 
the separate components corresponding to the solar ratio of 82 : 18 (Cameron 
\cite{cam}) and the r-process ratio of 54 : 46 (Arlandini et al. \cite{rs99}) 
are as follows:\\[1mm] 
\begin{tabular}{cccc}   
even-to-odd & 82 : 18 & \ \ \ \ & 54 : 46 \\
$\log gf\,(4554.000) =$ & $-1.011$ & & $-0.609$ \\
$\log gf\,(4554.034) =$ & $\,~~0.077$ & & $-0.097$\\
$\log gf\,(4554.052) =$ & $-0.790$ & & $-0.389$ \\ 
\end{tabular}\\[1mm]  
\noindent The absolute oscillator strength of $\lambda\,4554$, $\log gf = 0.162$, is 
taken from Wiese \& Martin (\cite{WM}). 

\begin{figure}
\resizebox{88mm}{!}{{\includegraphics{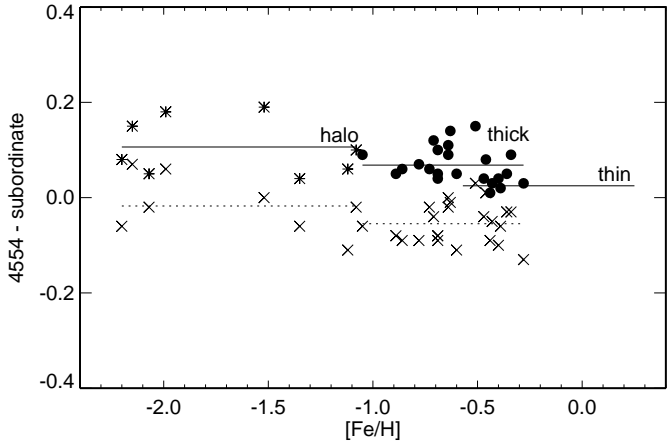}}} \vspace{-5mm} 
\caption[]{Comparison of barium abundances derived from the \ion{Ba}{ii}  
$\lambda\,4554$ and subordinate lines. The data obtained with the solar Ba isotope 
mixture are shown by filled circles (thick disk stars) and by asterisks (halo 
stars). The solid line for each stellar population indicates the mean value of 
abundance difference. Dotted lines correspond to similar values obtained with 
the r-process Ba isotope mixture and crosses for the thick disk and halo stars} 
\label{bars} 
\end{figure}
For all stars with both the resonance and subordinate \ion{Ba}{ii} lines 
available barium abundances were derived from $\lambda\,4554$ assuming a solar 
Ba isotope mixture. For the thin disk stars $\eps{}$\,(4554) coincides, on 
average, with $\eps{}$\,(subordinate lines), with the mean value of abundance 
difference equal to $0.02 \pm 0.04$ dex. As expected, the even-to-odd Ba isotope 
ratio in these stars is close to solar. For thick disk and halo stars values of 
$\eps{}$\,(4554) have been obtained assuming the r-process Ba isotope mixture, 
too. In Fig.~\ref{bars} the differences $\eps{}$\,(4554) $-\eps{}$\,(subordinate 
lines) are shown for both cases. At the even-to-odd Ba isotope ratio of 54 : 46 
the resonance and subordinate lines in the halo stars give, on average, the same 
abundances with the mean abundance difference of $-0.02 \pm 0.06$. This 
suggests a dominance of the r-process contribution to barium. For the thick disk 
stars this assumption results in underestimating $\eps{}$\,(4554) compared with 
$\eps{}$\,(subordinate lines) by $-0.05 \pm 0.04$ dex. For 3 representative thick 
disk stars, HD\,18757 ([Fe/H] = $-0.28$), HD\,3795 ([Fe/H] = $-0.64$) and 
HD\,22879 ([Fe/H] = $-0.86$) we have calculated $\eps{}$\,(4554) for a number of Ba 
isotope ratios. Using these results, i.e. $\eps{}$\,(4554) $-\eps{}$\,(subordinate 
lines) vs. even-to-odd Ba isotope ratio, we estimate that a fraction of the odd 
Ba isotopes of $\sim 35 \pm 10$\% should be adopted to put the abundances from 
different lines on a common scale. From this value the s-process contribution to 
Ba in thick disk stars is obtained as $30 \pm 30$\%.  

\section{Discussion} \label{discus}
\subsection{Abundance ratios [Eu/Ba] and a timescale for the formation of 
Galactic stellar populations}

\begin{figure}
\resizebox{88mm}{!}{{\includegraphics{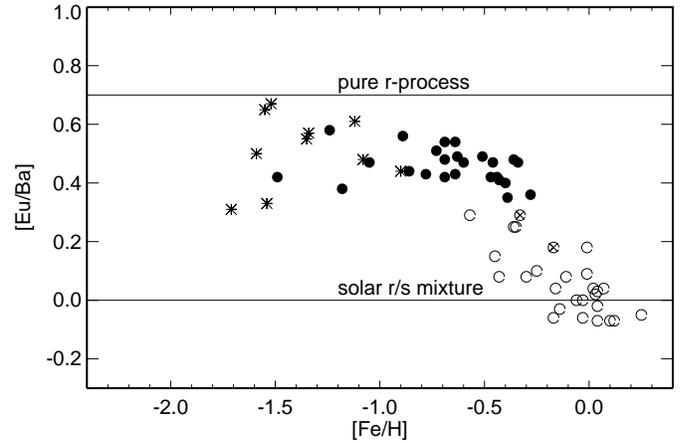}}}
\vspace{-5mm}
\caption[]{The run of [Eu/Ba] with [Fe/H]. 
Compare the values for the thick disk 
and thin disk stars in the region of overlapping metallicities. 
Symbols are the same as in Fig.\ref{eu_fe}}
\label{euba}
\end{figure}

[Eu/Ba] abundance ratios are shown in Fig. \ref{euba}. The solar abundance ratio 
of Eu to Ba contributed by the r-process (Arlandini et al. \cite{rs99}) relative 
to the total abundances, [Eu/Ba]$_r = 0.70$, is indicated in Fig. \ref{euba} by 
solid line. The new data confirm in general and improve in statistical sense the 
conclusions drawn in Paper~II; they also provide a fundament for new 
conclusions. 

Europium is significantly overabundant relative to barium in the halo and thick 
disk stars, and in the region of overlapping metallicities the [Eu/Ba] values do 
not reveal a clear distinction between these stellar populations. According to 
the deviation of observed values of [Eu/Ba] from [Eu/Ba]$_r = 0.70$ we can 
estimate the s-process contribution to barium. For the thick disk stars the 
[Eu/Ba] values are all between $0.35$ and $0.57$. This suggests that during the 
active phase of thick disk formation evolved low mass stars enriched the 
interstellar gas by s-nuclei of Ba with an s-process contribution to barium  
from 30\% to 50\%. This agrees with the estimate of the s/r-process ratio of 30 
: 70 ($\pm 30$\%) obtained above from our analyses of the even-to-odd Ba isotope 
ratios in the thick disk stars. According to the chemical evolution calculations 
of Travaglio et al. (\cite{eu99}) s-nuclei of Ba appear after about 0.5~Gyr from 
the beginning of the protogalactic collapse; in another $\sim 0.6$ Gyr the 
s/r-process ratio reaches 30 : 70, and in further 0.5 Gyr arrives at 50 : 50. 
Thus, {\it the thick disk population formed in the early Galaxy during an 
interval of $\sim$ 1.1 Gyr to 1.6 Gyr after the beginning of the protogalactic 
collapse}.  

For the halo stars the spread in the [Eu/Ba] values between $0.31$ and $0.67$ is 
rather large. The values close to [Eu/Ba]$_r = 0.70$ favour the dominance of the 
r-process in heavy element production; under such circumstances the star 
formation epoch is related to the first 0.5 Gyr after the beginning of the 
protogalactic collapse. The [Eu/Ba] values of $\sim 0.4$ give arguments for a 
duration of the halo formation of $\sim 1.5$ Gyr. Thus, both the oldest stars of 
the Galaxy and the ``late halo'' stars are found among the moderate 
metal-deficient halo stars (\,[Fe/H] $\geq -1.71$). This suggests that the 
metallicity of halo stars does not correlate with the stellar age. 

>From analyses of the [Fe/O] and [Mg/Fe] values Gratton et al. (\cite{Gratton00}) 
estimate that the formation of stars in {\it the halo and the thick disk} was 
fast (a few $10^8$ yr), i.e. shorter than the typical timescale of evolution for 
the progenitors of type Ia SNe. Our data on the [Eu/Ba] values suggest a longer 
duration of the {\it halo formation phase} of $\sim 1.5$ Gyr and a delay of the 
thick disk formation of about 1 Gyr. Such discrepancy may be connected with 
problems of the r-process and $\alpha$-process element abundances in metal-poor 
stars. We discuss these problems below. 

\subsection{Europium versus magnesium in halo stars and nucleosynthesis in the 
early Galaxy} \label{seumg}

\begin{figure}
\resizebox{88mm}{!}{{\includegraphics{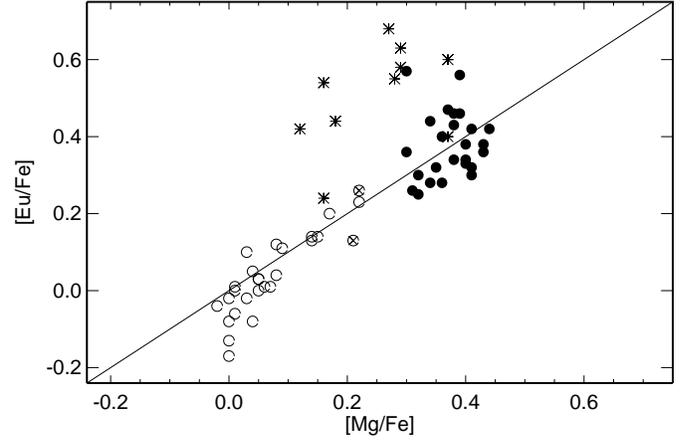}}}
\vspace{-5mm}
\caption[]{The run of [Eu/Fe] with [Mg/Fe]. 
Compare the values for the thick disk and halo stars. 
Symbols are the same as in Fig. \ref{eu_fe}}
\label{eumg}
\end{figure}

In Fig. \ref{eumg} [Eu/Fe] values are plotted against [Mg/Fe] values. As 
expected, the Eu abundance follows the Mg abundance in thin and thick disk 
stars. An exception is found in two thick disk stars, HD\,3795 and HD\,102200, 
that show an overabundance of Eu relative to Mg with [Eu/Mg] = $0.27$ and $0.23$, 
respectively. In 8 out of 10 halo stars europium is overabundant relative to 
magnesium with a mean value [Eu/Mg] = $0.31 \pm 0.06$. The observed values of 
[Eu/Mg] are independent of stellar metallicity (between $-0.90$ and $-1.59$), 
effective temperature (from 5110 K to 6130 K), surface gravity (from $3.12$ to 
$4.66$) and Mg enhancement (from $0.12$ to $0.37$). This gives reason to exclude an 
influence of possible methodical errors of the [Eu/Mg] value, connected with the 
treatment of NLTE effects or with the use of 1-D model atmospheres. The 
remaining two stars were discussed above as spectroscopic binary (HD\,99383) and 
a star with low Eu and Mg abundance. They have the [Eu/Mg] values close to 0. 
Thus, 5 halo stars added in this study confirm the observational finding 
reported earlier in respect to an overabundance of Eu relative to Mg in the 
halo. This gives strong evidence for different sites of Eu and Mg production. 

\begin{figure}
\resizebox{88mm}{!}{{\includegraphics{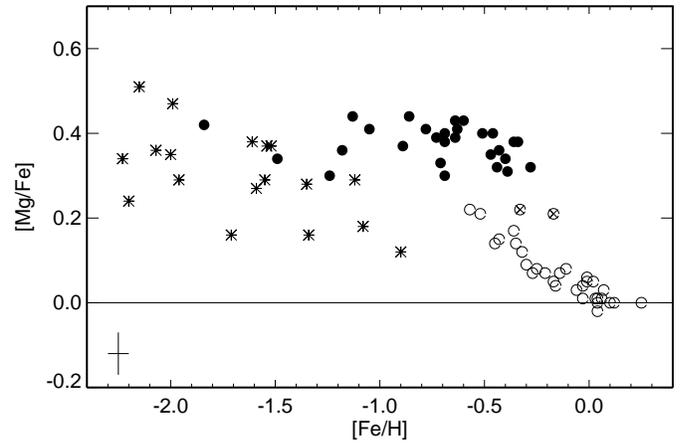}}}
\vspace{-5mm}
\caption[]{The run of [Mg/Fe] with [Fe/H]. Compare the spread of data for the 
thick disk and halo stars. Symbols are the 
same as in Fig. \ref{eu_fe}}
\label{mgfe}
\end{figure}

The following hypotheses could explain that overabundance. 

1.~~After the epoch of SN\,II dominance in nucleosynthesis additional production 
of Galactic magnesium occurred, and it was connected with stars of $M < 8 
M_\odot$. An argument for that can be found from inspection of the variation of 
the [Eu/Fe] and [Mg/Fe] values with overall metallicity for the thick disk 
stars. At [Fe/H] $> -1.0$ our data show a clear decline of the [Eu/Fe] abundance 
ratios with increasing metallicity of the order 
$${\rm [Eu/Fe]} =  0.22(\pm 0.04) - 0.25(\pm 0.07) {\rm [Fe/H]} \quad .$$ 
\noindent This supports the notion that during the thick disk formation phase 
iron starts to be produced in SNe\,I and its production rate is higher than that 
for Eu. If SNe\,II are the main source of Galactic magnesium, a similar decline 
is expected for the [Mg/Fe] values. 
 In Fig. \ref{mgfe} the run of [Mg/Fe] with [Fe/H] is shown. 
For our sample of 26 thick disk stars the 
[Mg/Fe] values taken mostly from the work of Fuhrmann (\cite{Fuhr3, Fuhr00}) are 
all between $0.30$ and $0.44$, and only a marginal decline of these ratios with 
increasing metal abundance can be detected. 
 About 20  thick disk stars of the Gratton et al. sample (\cite{Gratton00}, 
their  Fig. 4)  show [Mg/Fe] values between $0.24$ and $0.44$, and there is only 
a hint of a small decline of this ratio with increasing metallicity. 
Prochaska et al. (\cite{proch}) 
suggest a small decline of the [$\alpha$/Fe] values for their sample of 10 thick 
disk stars. Thus, the production rate of Mg during the thick disk formation 
phase seems to be higher compared with that for Eu. 

Another way to check the hypothesis of an additional source of Galactic 
magnesium is to compare Eu abundances with oxygen abundances because one 
commonly believes that O is mostly produced in SNe\,II. However, the data on the 
[O/Fe] values available in the literature are confusing (see, for example, the 
data collected by Melendez et al., \cite{o_ab}, their Table 10). Nissen et al. 
(\cite{nis02}) discuss in detail possible sources of inconsistencies in O 
abundances determined from the different methods and show that careful analyses 
of the ultraviolet OH, the forbidden [\ion{O}{i}] $\lambda\,6300$ lines and the 
\ion{O}{i} triplet at 7774 \AA \ using 1-D model atmospheres give consistent O 
abundances for metal-poor dwarfs and subgiants. For two stars in common, 
HD\,97320 and HD\,298986, we use their data on the [O/Fe] values based on the 
[\ion{O}{i}] $\lambda\,6300$ line and 1-D model atmospheres, [O/Fe] = $0.33$ and 
$0.38$, respectively. Differences in adopted model atmosphere parameters are 
small. $\Delta \Teff$ (this work -- Nissen et al., \cite{nis02}) = 130 K and 60 
K, $\Delta \log g = 0.11$ and $0.09$, $\Delta$[Fe/H] = $0.03$ and $0.01$, 
respectively. We obtain [Eu/O] = $0.07$ for HD\,97320 and [Eu/O] = $0.16$ for 
HD\,298986. When a correction of $-0.10$ dex (from Table 6 of Nissen et al., 
\cite{nis02}) is applied accounting for the influence of granulation on the O 
abundance, the [Eu/O] values become larger, $0.17$ and $0.26$. Both the 
[\ion{O}{i}] $\lambda\,6300$ and \ion{Eu}{ii} $\lambda\,4129$ lines originate 
from the ground state of the dominant ionization stage of the corresponding 
atoms, and their intensities are essentially insensitive to temperature 
variations in a 3-D model atmosphere. However, Nissen et al. (\cite{nis02}) 
point out that the effect of 3-D granulation on the [\ion{O}{i}] line comes from 
differences in the continuous opacities between 3-D and 1-D models. This effect 
should be much smaller for the \ion{Eu}{ii} line because it is much stronger 
compared with the [\ion{O}{i}] line. 

Oxygen abundances based on the [\ion{O}{i}] $\lambda\,6300$ line are found in 
the literature also for the halo star HD\,103095. Israelian et al. 
(\cite{is_o}), Balachandran \& Carney (\cite{ba_o}) and Spite \& Spite 
(\cite{ss_o}) give [O/Fe] = $0.37$, $0.33$ and $0.48$, respectively. Using a 
mean value [O/Fe] = $0.39$ we arrive at [Eu/O] = $0.16$. Thus, at least, in 2 
halo stars europium is overabundant not only relative to magnesium but also 
relative to oxygen. This does not exclude with certainty the existence of an 
additional source of Galactic magnesium. 

2.~~Current theoretical models of the r-process are valid, but Eu is mostly 
produced in low-mass SNe\,II while Mg is synthesized in larger amounts in the 
higher mass stars. However, mixing of the interstellar matter was insufficient 
in the early Galaxy up to the epoch with [Fe/H] $\sim -1.0$. In this case there 
should be stars with an overabundance of Mg relative to Eu, which were born near 
exploded high-mass stars, and stars with [Eu/Mg] $> 0.0$ born near low-mass 
SNe\,II. Due to an unknown selection effect we observe stars with high Eu 
abundance (resulting in [Eu/Mg] $> 0.0$) and do not observe stars with low Eu 
abundance. 

Out of 6 halo stars with no europium abundance available two stars, HD\,84937 
([Fe/H] = $-2.07$, [Mg/Fe] = $0.36$) and BD$-4^\circ 3208$ ([Fe/H] = $-2.23$, 
[Mg/Fe] = $0.34$), could be candidates for the ``high Mg/low Eu abundance'' 
sample. Their \ion{Eu}{ii} $\lambda\,4129$ line can not be extracted from noise, 
and the upper limit of the [Eu/Fe] value is estimated as 0.4 dex. Two stars, 
BD$+66^\circ 268$ ([Fe/H] = $-2.20$, [Mg/Fe] = $0.24$) and BD$+34^\circ 2476$ 
([Fe/H] = $-1.96$, [Mg/Fe] = $0.29$), are more likely ``low Mg/high Eu 
abundance'' stars because for both of them the upper limit of the [Eu/Fe] value 
is higher (0.5 dex) and the [Mg/Fe] value is lower. For the remaining two halo 
stars spectra of $\lambda\,4129$ have not been observed. Thus, in total, 10 halo 
stars of our sample can be related to ``low Mg/high Eu abundance'' stars and 4 
stars to a ``high Mg/low Eu abundance'' sample. 

The large spread of the [Mg/Fe] values in the halo stars 
 seen in Fig. \ref{mgfe} can serve as an 
argument for this hypothesis. The 16 halo stars of our sample have [Mg/Fe] 
ratios between $0.12$ and $0.51$ with no correlation to metallicity. Such a 
spread can not be caused by errors of Mg abundance determinations because in the 
region of overlapping metallicities the thick disk stars show a much narrower 
range of [Mg/Fe] values of $0.30$ to $0.44$. 

We have inspected the recent work of Fulbright (\cite{Fulb00}), where Eu and Mg 
abundances are presented for stars in the metallicity range covered in this 
study. Fulbright does not identify the membership of individual stars in 
particular stellar populations of the Galaxy, and we have selected from his 
sample 17 halo stars with $\log g \geq 3.0$, based on their metallicities 
([Fe/H] $\leq -0.9$). Seven of the stars with relatively low Mg abundance (the 
[Mg/Fe] value is between $0.06$ and $0.29$) reveal overabundances of Eu relative 
to Mg with a mean value [Eu/Mg] = $0.39 \pm 0.12$ while the remaining 10 stars 
with [Mg/Fe] between $0.35$ and $0.55$ have the [Eu/Mg] values from $-0.38$ to 
$0.18$ with the mean value [Eu/Mg] = $-0.06 \pm 0.21$. Both samples of stars 
have the same range of metallicity, i.e. from $-0.92$ to $-1.60$. Thus, a 
hypothesis of \emph{insufficient mixing} of the interstellar matter during the 
halo formation phase seems to be reasonable. 

\section{Concluding remarks}

In the literature moderately metal-deficient halo stars with [Fe/H] from $-1.5$ 
to $-1.0$ are often referred as ``late halo'' stars and attention is mainly 
focused on extremely metal-poor stars when problems of the early Galaxy are 
studied. In this paper we have demonstrated that moderate metal-deficient stars 
give useful information about nucleosynthesis in the Galaxy (the s/r-process 
ratio, different sites of Mg and Eu production), about mixing of the 
interstellar matter, and about the timescale for the \emph{formation} of the 
Galactic halo and thick disk. 

After the work of Gilmore \& Reid (\cite{gilmore}) who offered the evidence for 
the existence of the thick disk stellar population\footnote {Marsakov \& Suchkov 
(1977) were probably the first who suggested the existence of the ``intermediate 
halo'' stellar population with the kinematics and metallicity corresponding to 
the stellar population now named ``thick disk''} the key question is how it is 
related to the thin disk and halo in terms of the Galaxy's chemical and 
dynamical evolution. Clear evidence of chemical distinction of the thick from 
thin disk was given by Gratton et al. (\cite{Gratton96}) and Fuhrmann 
(\cite{Fuhr3}) from analyses of the [O/Fe] and [Mg/Fe] abundance ratios and by 
our previous studies (Papers~I and II) from analyses of the [Eu/Ba] values. A 
step-like decrease in $\alpha$/Fe and Eu/Ba ratios at the thick to thin disk 
transition indicates a phase of nearly ceased star formation before the earliest 
stars of the thin disk developed. In this paper we have compared chemical 
properties of the thick disk and halo stars, and it is important that they have, 
in part, overlapping metallicities. Surprisingly good correlations of various 
chemical elements found in the thick disk stars ([Mg/Fe] between $0.30$ and 
$0.44$; [Eu/Ba] between $0.35$ and $0.57$ with the hint of a decline with 
increasing metal abundance; a clear decline of the [Eu/Fe] values with 
increasing metal abundance) suggest that the thick disk stellar population 
formed from well mixed gas during a short time interval of $\sim 0.5$ Gyr, 
according to our estimate. 
We note that strong support to the present conclusion that the thick disk is a 
homogeneous, chemically old population comes from the results of Nissen \& 
Schuster (\cite{nis97}), Fuhrmann (\cite{Fuhr3}) and Gratton et al. 
(\cite{Gratton00}); they found that the thick disk stars have low [Fe/Mg] and 
[Fe/O] values (equal or even lower than those found for halo stars of similar 
metal abundance), with very small intrinsic scatter. 
In opposite, the large spread in element abundance ratios found in 
halo stars with metal abundances up to [Fe/H] = $-0.9$ points at insufficient 
mixing of the interstellar matter during the halo formation phase.
Most probably,  metal abundance did not correlate with time during the
evolution of the Galactic  halo. All the thick disk stars investigated here
reveal an s-process element  enrichment with a fraction of the s-process
contribution to barium from 30\% to  50\%. We conclude that the thick disk
stellar population must have formed on a  timescale of between 1.1 and 1.6 Gyr
after the beginning of the protogalactic  collapse. In this paper we study
only a few ``metal-weak thick disk'' stars. It  would be important to extend a
sample of such stars. Among the moderately  metal-deficient halo stars we see
also an enrichment of s-process elements. This  suggests that the phases of
halo and thick disk formation overlapped in time. 

Further theoretical and observational work is required to understand the problem 
of the Eu overabundance relative to Mg in halo stars. It would be important to 
detect not only Eu/Mg but also Eu/$\alpha$ abundance ratios in halo stars and to 
extend the sample of stars both in number and metallicity. 

\begin{acknowledgements}
ML acknowledges with gratitude the Max-Planck Institut of Astrophysics for the 
partial support of this study and the Institute of Astronomy and Astrophysics of 
Munich University for warm hospitality during a productive stay in December 2001 
- February 2002. We thank Klaus Fuhrmann for providing reduced spectra and 
stellar parameters of the three stars before publication, for fruitful 
discussions. We are grateful to Wolfgang Hillebrandt, Roberto Gallino, Andreas
 Korn, Johannes 
Reetz and Vladimir Marsakov for valuable help and useful discussions. ML and 
BT have  
been partially supported by the Russian Basic Researches Fund (grants  
02-02-17174 and 02-02-06911, correspondingly). 
\end{acknowledgements}

\end{document}